# Difference of Oxide Hetero-Structure Junctions with Semiconductor Electronic Devices

XIONG Guang-Cheng, CHEN Yuan-Sha, CHEN Li-Ping, LIAN Gui-Jun

Department of Physics

Peking University, 100871 Beijing, P. R. China

Charge carrier injection performed in $Pr_{0.7}Ca_{0.3}MnO_3$ (PCMO) hetero-structure junctions exhibits stable without electric fields and dramatic changes in both resistances and interface barriers, which are entirely different from behaviors of semiconductor devices. Disappearance and reversion of interface barriers suggest that the adjustable resistance switching of such hetero-structure oxide devices should associate with motion of charge carriers across interfaces. The results suggested that injected carriers should be still staying in devices and resulted in changes in properties, which guided to a carrier self-trapping and releasing picture in strongly correlated electronic framework. Observations in PCMO and oxygen deficient $CeO_{2-\delta}$ devices show that oxides as functional materials could be used in microelectronics with some novel properties, in which interface is very important.

Development of devices in microelectronics has induced remarkable changes in daily life for our modern society. Main structure of the semiconductor electronic devices is simple double layers or sandwiches using one type of semiconductor connecting with another that called as heterostructure consist of different materials. Heterostructure results in diodes and transistors that are operated by carrier injection in electric fields. In semiconductor physics, conductivity of semiconductors can be adjusted by doping various impurities, which leads to different kinds of semiconductor devices. Oxides are usually used as insulating materials, for example, as gate dielectric materials in microelectronics industry.

After the discovery of high-temperature superconductors several unexpected properties, such as colossal magnetoresistance in doped manganites, were fond in oxides. Different with semiconductors, doped manganites belong to strongly correlated materials with strongly correlated electrons that exhibit variation of properties in electric fields. For example, resistance of doped manganite single crystals presents a drastic drop more than three orders of magnitude at 20 K in electric fields[1]. Several groups also reported influence of electric currents on transport properties of thin films[2-4]. In $Pr_{0.7}Ca_{0.3}MnO_3$ and other poor-conductive oxides, resistance of samples could switch from high resistance states (HRS) to low resistance states (LRS) simply by applying electric fields, which may be used as nonvolatile resistance-memory[5-8]. Facing such complex oxides, combination of oxides as functional materials in devices may lead to some fantastic phenomena, in which comparison with semiconductor electronic devices would be interesting and important for getting information about such novel oxide electronics devices.

Double layer diode with rectifying behaviors is basic structure in semiconductor devices that presents low resistance in forward condition and high resistance in reverse, which exhibits influence of interface barriers. Injection of charge carriers across interface by electric fields leads to some novel results, such as light emitting in light-emitting diodes (LED). Turning off electric fields for LED, emitted light disappears and devices return back to original states. If carrier injection was performed in transition-metal-oxide junctions, influence on samples can be easily examined by current-voltage (*I-V*) measurement. Influence of carrier injection for transition metal oxide junctions would be interesting and also enclose information for understanding oxide electronics devices.

By using pulsed laser deposition technique as previously described[9], oxides such as $Pr_{0.7}Ca_{0.3}MnO_3$ (PCMO), $SrTiO_3$ (STO) and $CeO_2$ can be easily deposited on substrates to fabricate layered devices. Because Nb-doped $SrTiO_3$ (Nb-STO) is *n*-type oxide semiconductor[10, 11], it is easy to fabricate oxide junctions by using Nb-STO as substrates. Resistance of Nb(0.6%)-STO (with Nb-doping concentration of 0.6 wt.%) substrates was about 0.3 $\Omega$ at room temperature and decreased to 0.03 $\Omega$ at 20 K. In experiments, device properties were examined by *I-V* measurement in current scan cycles as $0 \rightarrow + I_{MAX} \rightarrow 0 \rightarrow - I_{MAX} \rightarrow 0$. In measured data, linear *I-V* data show contribution of constant resistance and rectifying curves exhibit influence of interface barriers. In plots, numbers indicate the measurement sequence and arrows point direction of current sweeping.

Using insulating STO buffer layers with geometry as indicated in the inset of Fig. 1A, PCMO/Nb-STO devices were fabricated that have equivalent circuit simple as resistance of a PCMO layer connecting in series with a diode [12]. Figure 1A shows *I-V* data of a heterostructure PCMO/Nb(0.6%)-STO device with the special geometry measured at room temperature, in which the PCMO film on STO buffer layer was about 1 mm with resistance of about 5-7 k$\Omega$. In low voltage region (V < 0.1 V), resistance of the device was higher than 100 kΩ. The rectifying behavior obtained in this device demonstrates the formation of interface barriers as heterostructure junctions that observed in normal semiconductor diodes. In forward condition device resulted in low resistance about 10 kΩ (for 0.1 mA with 1 V).

Figure 1B shows *I-V* data of the PCMO/Nb-STO device measured with higher currents also at room temperature. It is interesting that resistance of the PCMO/Nb-STO device changed from HRS to LRS of 100 Ω in positive-going current scan. The LRS switched by electric fields was still stable with nearly linear character without electric fields. When restarting the next negative current scan, resistance of the device returned back to HRS with an abruptly jump in some negative current step. The process, switching from HRS to LRS in positive-current scan and resetting from LRS to HRS in negative-current scan, as a device function can be repeated as exhibited in Fig. 1B, and that this should respond to carrier injection and release for such oxide device. The *I-V* data in electric fields demonstrates resistance-switching behavior from HRS about 100 kΩ (for 0.001 mA) to LRS about 100 Ω.

Obviously, the nearly linear character of LRS should be influence of carrier injection obtained in the PCMO/Nb-STO device. In order to confirm behavior of injected carriers, another PCMO/Nb-STO junction was fabricated as simply deposited PCMO layer on a Nb(0.6%)-STO substrate. For examining contribution of thermal diffusion of defects and oxygen vacancies, carrier injection measurement was performed at low temperature. Figure 2 shows *I-V* data of the PCMO/Nb(0.6%)-STO junction measured at 80 K with two current scan cycles. When cooling down to low temperature, effective resistance of the PCMO/Nb(0.6%)-STO junction (measured by $5 \times 10^{-4}$ mA) increased from 43 kΩ (at room temperature) to 700 kΩ (at 80 K). In a positive ($I_{\text{Maximum}}$ = 1 mA) current scan, effective resistance of the device dropped from HRS of 700 kΩ to LRS of about 200 Ω with several step jumps that shown changes of over three orders of magnitude. The LRS was also stable with nearly linear character after removing the electric field unless restarting next negative current scan. In a negative-current scan with increasing current, the stable setting LRS returned back to HRS with abruptly jump. The process can be repeated as device function as exhibited in Fig. 2. Because measurement was performed at low temperature, thermal diffusion should not be a reason for such switching behavior in oxide junctions.

In semiconductor LED junctions, carriers are injected in forward condition that the nonequilibrium charge carriers will recombine and lead to light emission. In PCMO/Nb-STO junctions, the dramatic changes of transport properties in forward condition demonstrate the influence of carrier injection. It is interesting that the switched LRS is stable even without electric field, which is not as same as that observed in semiconductor LED junctions. It should be pointed out that the nearly linear *I-V* character of LRS exhibits behavior of typical pure resistance, which indicates disappearance of interface barriers. As influence of carrier injection, therefore, LRS reveals dramatic changes in both resistances of PCMO and interface barriers. The disappearance and reversion of interface barriers at 80 K should be evidence that this process relates to motions of carriers crossing interface. The stable switching states indicate behavior of the oxide electronic devices operated by carrier injection that exhibit large difference with semiconductor devices.

Because doped manganites belong to the strongly correlated materials that are different with semiconductors, therefore, obtained novel phenomena in these devices should be natural. In the process of carrier injection, once a charge carrier climbed over an interface barrier, the carrier can fell on a metastable state as that happed in semiconductor LED. The injected carriers in metastable states, however, would bring influence on effective ion valence and then cause changes on effective ion radius. Due to the strong electron-phonon interactions in doped manganite, the injected carrier with self-caused effects could be self-trapping in metastable states and result in changes in electronic band structure that guided to a carrier self-trapping picture.

By using oxygen-deficient $CeO_{2-\delta}$ we further examined influence of carrier injection in other transition-metal-oxide devices. Being as an insulating material, cerium dioxide of $CeO_2$ with high thermal stability has been extensively studies, for example, as gate dielectric material for semiconductor devices. Oxygen vacancy in $CeO_2$ results in appearance of $Ce^{3+}$ and conductivity. We grew oxygen-deficient $CeO_{2-\delta}$ layers on low-resistance $La_{0.7}Sr_{0.3}MnO_3$ (LSMO) films and measured the *I-V* data. Figure 3 shows room-temperature *I-V* data of an LSMO/$CeO_{2-\delta}$ device. When using a probe on LSMO layer as positive electrode, rectifying behavior of the LSMO/$CeO_{2-\delta}$ device was similar to that obtained in PCMO/Nb-STO junctions. This result indicated formation of barriers in the LSMO/$CeO_{2-\delta}$ interfaces and that the oxygen-deficient $CeO_{2-\delta}$ acted as *n-type* oxide semiconductors. As observed in PCMO/Nb-STO devices, stable resistance-switching behavior for carrier injection was obtained in the LSMO/$CeO_{2-\delta}$ device that resistance of the device decreased from HRS about 100 kΩ to LRS about 1 kΩ in positive-current scans and reset in negative-current scans, which associated with changes in the interface barriers. Results in the LSMO/$CeO_{2-\delta}$ device exhibit formation and changes of interface barriers that may be a reason of resistance switching observed in such broad poor-conductive oxide devices.

In semiconductor, carrier injection across interface is important for developing several devices such as light-emitting and semiconductor laser. In transition-metal-oxide devices, carrier injection in electric fields exhibited dramatic changes in both resistances and interface barriers. The observations in PCMO and $CeO_{2-\delta}$ devices with stable switching states when removing the electric fields demonstrate difference with semiconductor devices.

Transition metal oxides such as doped manganites have strongly correlated electrons that are different with semiconductors. The obtained setting and resetting situations in broad oxide devices[1-8] suggest that oxide used as functional materials in microelectronics should results in fantastic and unexpected properties, especially considering variation in the complex oxides. Our experiments exhibit interesting results of carrier injection in oxide electronic devices that related with dramatic changes in both resistances and interface barriers. The observation in strongly correlated electronic framework guided to a carrier self-trapping picture. The oxygen-deficient $CeO_{2-\delta}$ devices demonstrated the importance of interface for such novel oxide electronics devices.

## Acknowledgements

We thank C. Y. Fong, Z. Z. Gan, Z. X. Zhao and Y. H. Zhang for helpful discussions. We wish to thank the National Natural Science Foundation of China for financial support.

References

1. Asamitsu A, Tomioka Y, Kuwahara H and Tokura Y 1997 Nature **388** 50

2. Hu F X and Gao J 2004 Phys. Rev. B **69** 212413

3. Masuno A, Terashima T, Shimakawa Y and Takano M 2004 Appl. Phys. Lett. **85** 6194.

4. Jain H, Raychaudhuri A K, Mukovskii Ya M and Shulyatev D 2006 Appl. Phys. Lett. **89** 152116

5. Liu S Q, Wu N J and Ignatiev A 2000 Appl. Phys. Lett. **76** 2749

6. Beck A, Bednorz J G, Gerber Ch, Rossel C and Widmer D 2000 Appl. Phys. Lett. **77** 139

7. Nakamura A, Matsunaga K, Tohma J, Yamamoto T and Ikuhara Y 2003 Nature Materials **2** 453

8. Kim D S, Kim Y H, Lee C E and Kim Y T 2006 Phys. Rev. B **74** 174430

9. Song X F, Lian G J and Xiong G C 2005 Phys. Rev. B **71** 214427

10. Carles A L, Rogers T, Proce J C, Rudman D A and Herman R 1998 Appl. Phys. Lett. **72** 3065

11. Yoshida A, Tamura H, Gotoh K, Takauchi H and Hasuo S 1991 J. Appl. Phys. **70** 4976

12. Indium was used as electrodes. For In/PCMO interface, contact resistance is less than 1 Ω with linear I-V behavior. I and V electrodes on Nb-STO were separated to exclude the signal of Schottky-barrier from In/Nb-STO interface.

**Figure Captions**

Figure 1 (color). *I-V* data measured at room temperature for a PCMO/Nb-STO junction with special geometry in low-current region A and high-current region B. Inset of Fig. 1(a) is schematic drawing of the special geometry with insulating STO as buffer layers.

Figure 2 (color). *I-V* data of a PCMO/Nb-STO junction measured at 80 K. Inset is schematic drawing of the junction with electrode geometry that PCMO is as positive.

Figure 3 (color). *I-V* data measured at room temperature of LSMO/$X_{\varepsilon O2-\delta}$ device. Inset is schematic drawing of the electrode geometry that the electrode on LSMO is positive.

**Figures**

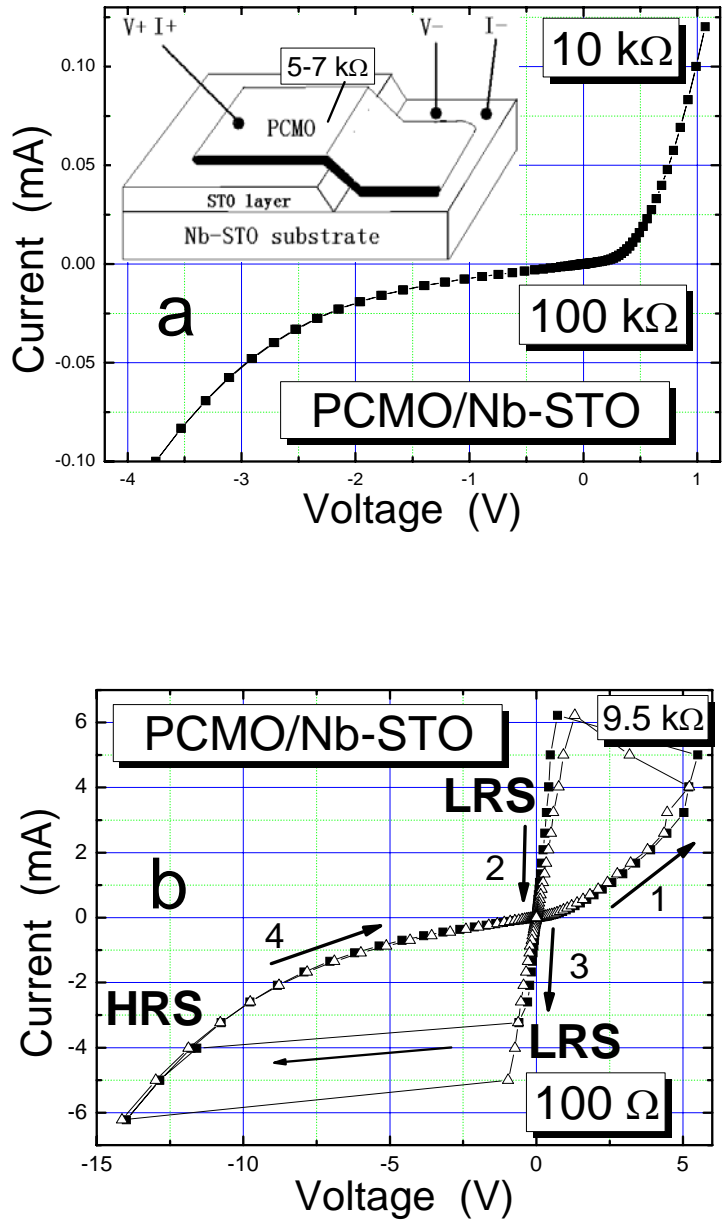

Figure 1

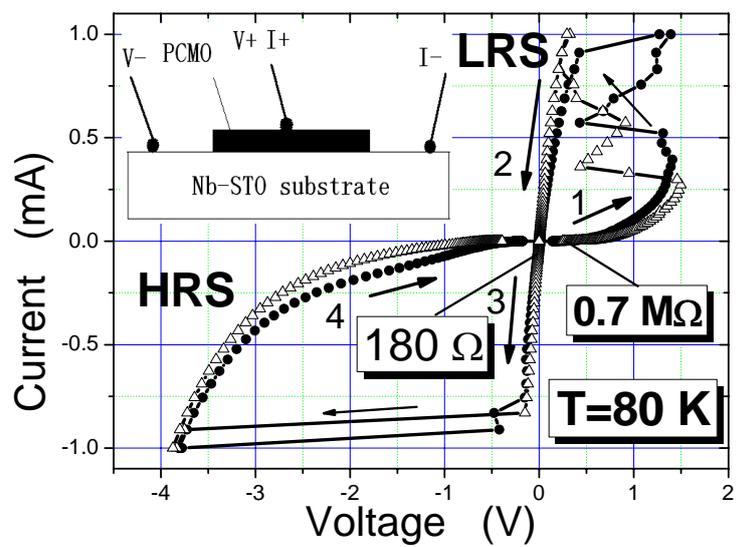

Figure 2

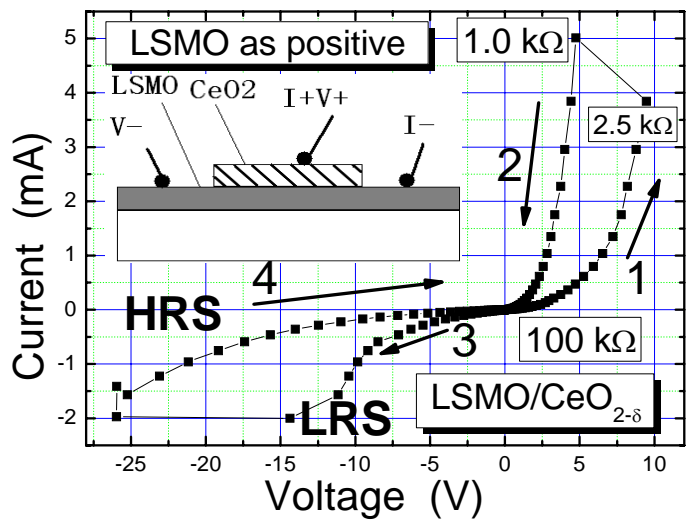

Figure 3